\documentclass[nohyper,11pt,letterpaper]{JHEP3}
\usepackage[dvips]{epsfig}
\usepackage{amsmath}
\newcommand{\lf}{\left [}
\newcommand{\rf}{\right ]}
\newcommand{\be}{\begin{equation}}
\newcommand{\ee}{\end{equation}}
\newcommand{\bea}{\begin{eqnarray}}
\newcommand{\eea}{\end{eqnarray}}
\newcommand{\ba}{\begin{eqnarray}}
\newcommand{\ea}{\end{eqnarray}}
\newcommand{\nn}{\nonumber \\}

\newcommand{\beq}{\begin{equation}}
\newcommand{\eeq}{\end{equation}}
\newcommand{\beqa}{\begin{eqnarray}}
\newcommand{\eeqa}{\end{eqnarray}}
\newcommand{\beqar}{\begin{eqnarray*}}
\newcommand{\eeqar}{\end{eqnarray*}}

\def\t6 {T_\mt{D6}}

\newcommand{\mt}[1]{\textrm{\tiny #1}}

\title{(Un)attractor black holes in higher derivative AdS gravity}
\author{Dumitru Astefanesei,$^a$ Nabamita Banerjee,$^b$ and Suvankar Dutta$^b$\\

$^a$Max-Planck-Institut f\"ur Gravitationsphysik,
Albert-Einstein-Institut, 14476 Golm, Germany\\
$^b$ Harish-Chandra Research Institute, Chhatnag Road, Jhusi,
  Allahabad 211 019, India\\

\\E-mail: \email{dumitru@aei.mpg.de, nabamita,suvankar@hri.res.in}}

\abstract{ We investigate five-dimensional static (non-)extremal 
black hole solutions in higher derivative Anti-de Sitter gravity 
theories with neutral scalars non-minimally coupled to gauge fields. 
We explicitly identify the boundary counterterms to regularize the 
gravitational action and the stress tensor. We illustrate these 
results by applying the method of holographic renormalization to 
computing thermodynamical properties in several concrete examples. 
We also construct numerical extremal black hole solutions and discuss 
the attractor mechanism by using the entropy function formalism.}


\preprint{AEI-2008-037}

\begin{document}{\vskip 1cm}


\section{Introduction}
Among the fundamental interactions, gravity is very special. Gravity
couples via a dimensional coupling constant, the Newton constant
$G_N$, and so it is intrinsically non-renormalizable. Irrespective of
the fundamental nature of quantum gravity, the gravitational
low-energy degrees of freedom are encoded in the metric of spacetime
itself. However, there is no reason to believe that the effects of our
present theory are the whole story at the highest energies.  Indeed,
non-renormalizability can be interpreted as a natural feature of a
theory for which the action is not fundamental but arises as an
effective action in some energy limit. At high enough energies --- for
sufficiently strong curvatures and sufficiently small distances ---
new interactions and new degrees of freedom will be required.

The fact that the gravitational action is proportional to {\it R and
  only R} is not due to any symmetry and, unlike other theories, can
not be argued on the basis of renormalizability.  Indeed, the low
energy effective gravity action that obeys principle of equivalence
and general covariance has a generic structure. That is the usual
Einstein action plus a series of all possible interactions which are
consistent with general covariance and local Lorentz invariance, i.e.,
higher curvature terms and also higher derivative terms involving the
`low-energy' matter fields.

The effects of heavy particles appear to be local interactions when
viewed at low energy. That is the fields at different spacetime points
are independent degrees of freedom with independent quantum
fluctuations.  One important caveat related to the interpretation of
gravity as a (local) effective field theory is as follows: in a local
field theory one expects an entropy proportional to the volume, but
that is not true for black holes.  In classical gravity, a fixed
energy-density in a sufficiently large volume will collapse into a
black hole.\footnote{In quantum gravity, the existence of local
  operators is problematic due to the causality. It is well known that
  the commutator of space-like separated local operators should be
  zero. However, since the gravity is {\it dynamical} the metric
  itself fluctuates and so the space-like intervals are not well
  defined.} However, the holographic principle 
  \cite{'t Hooft:1993gx} was proposed to rescue
this situation: gravity in $D$ dimensions is equivalent with a local
field theory in $D-1$ dimensions. The AdS/CFT correspondence 
\cite{Maldacena:1997re} (see \cite{Nastase:2007kj} for a nice recent 
set of lectures) is 
a concrete realization of the holographic principle. Such correspondence
is referred to as duality in the sense that the supergravity (closed
string) description of D-branes and the field theory (open string)
description are different formulations of the same physics. This way,
the infrared (IR) divergences of quantum gravity in the bulk are
equivalent to ultraviolet (UV) divergences of dual field theory living
on the boundary. When we specify the CFT and say on which space it
lives we are implicitly providing a set of counterterms for the
gravity solution. These counterterms are local and depend only on the
intrinsic boundary geometry \cite{Henningson:1998gx,Balasubramanian:1999re}
(see, also, the reviews \cite{Skenderis:2000in}) --- a different method 
was proposed in \cite{Kofinas:2006hr}.  
Thus, one can compute the thermodynamical quantities in the gravitational 
side by employing the quasilocal formalism of Brown and York 
\cite{Brown:1992br} supplemented by the boundary 
counterterms. The connection between the holographic charges 
and the various alternative definitions of conserved charges in AdS was 
explored in \cite{Hollands:2005wt}.

In studying string theories at low energy scales, the massive states
may be integrated out to yield an effective action for the massless
modes, with the same symmetries as the original string theory. Thus
while the (super)gravity action is unique if we restrict to terms with
two derivatives, interactions quadratic or higher order in the
curvature tensor are allowed by the symmetries and so appear as well.
However, such terms will require a {\it dimensional constant} to
appear along with the derivatives. In string theory this constant
turns out to be $\alpha'$, the inverse string tension.\footnote{This
constant defines what is meant by `slowly varying fields' in the
sense that the derivative corrections may be ignored for fields that
are slowly varying on the scale of the string length $l_s\sim
  \sqrt{\alpha'}$.}

In this paper we investigate charged AdS black holes in the presence 
of higher derivative terms. We must note that, unlike in 
general relativity, in the presence of higher derivative corrections 
there are two families of solutions. We propose counterterms that 
regularize the action and the stress tensor of both branches (for 
horizons with spherical, toroidal, and hyperbolic topologies).

We obtain the stress tensor and the conserved charges for {\it exact} 
static charged non-extremal black hole solutions with Gauss-Bonet (GB) 
term and find perfect agreement with Wald formalism \cite{Wald:1993nt}. 
In the extremal limit we explicitly show that the near horizon geometry 
of the solution remains $AdS_2\times S^3$ after including 
$\alpha'$-corrections. The results we obtain provide a robust check 
of the entropy function formalism \cite{Sen:2005wa}. Indeed, 
we find that, for our {\it exact} solutions, 
the radius of $AdS_2$ receives corrections but the near horizon 
geometry remains $AdS_2\times S^3(H^3)$. In this way we obtain the 
generalization of Bertotti-Robinson geometries 
\cite{Bertotti:1959pf} with GB term. 

We also apply the counterterm method to $5$-dimensional charged black 
hole solutions in gravity theories with $U(1)$ gauge fields 
and neutral scalars. We obtain numerical solutions and generalize 
the results of \cite{Liu:2004it} by including the higher derivative terms. 

In the extremal limit we study the attractor mechanism by using the 
entropy function formalism \cite{Sen:2005wa,Astefanesei:2006dd}. 
This method is based on the near horizon geometry 
and its enhanced symmetries but does not provide a proof for the 
existence of a complete solution in the bulk. For some special values 
of the couplings, we present numerical solutions with a finite 
horizon --- this confirms the results in \cite{Astefanesei:2007vh} 
where the equations of motion in the bulk were solved perturbatively 
order by order. Thus, we can safely apply the entropy function formalism. 

An overview of the paper is as follows: in section 2 we study in detail 
the AdS charged black holes with GB term. We compute the stress 
tensor and the conserved charges of the exact non-extremal black hole 
solution by using the counterterm method and compare with the results 
obtained by Wald formalism. We present a preliminary 
discussion on the thermodynamics in both, canonical and grand-canonical 
ensembles. We also study the extremal limit and interpret our results 
within the entropy function formalism.
Section 3 is dedicated to studying black hole solutions in AdS gravity 
with U(1) gauge fields non-minimally coupled 
to scalars in the presence of GB term. We present numerical 
non-extremal solutions and discuss in detail their properties by using the 
counterterms proposed in section 2. In section 4 we study the extremal limit 
in the case of massless scalar fields, construct numerical solutions, and 
investigate the attractor mechanism for these solutions. In section 5, we 
discuss our results. An appendix gives some calculational details on Wald 
formalism.


\section{Charged AdS black holes with Gauss-Bonet term}
In this section we compute the conserved charges of AdS charged black 
holes with GB term by using both, the counterterm method and Wald formalism. 
The GB term is a very natural correction term to the Einstein action in 
the sense that the equations of motion contain no more than second 
derivatives in time. The main reason we are interested in GB term is due 
to the existence of {\it exact} solutions. In the extremal limit, we explictly 
check that the near horizon geometry still remains $AdS_2\times S^3(H^3)$. We 
also use the entropy function formalism to interpret our results.

\subsection{Non-extremal case}
\label{section21}
AdS spacetime is a maximally symmetric Lorentzian space ($i.e.$ the number
of Killing vector fields is the same as for flat spacetime) with constant
negative curvature --- in $D$-dimensions the symmetry group is $SO(D-1,2)$ 
and the topology is $AdS_D\equiv R^{D-1}\times S^1$.
AdS spacetime arises as the natural ground state of gauged supergravity 
theories and plays an important role in understanding holography in string 
theory.

As we are interested in AdS gravity with higher derivatives (see \cite{Garraffo:2008hu} 
for a recent review), we begin by 
establishing our conventions for the action. Exact solutions are presented 
in \cite{torii-maeda}. Note, however, that our conventions differ from the ones 
in \cite{torii-maeda} and we also correct some important typos.

In this section, we will focus on a 5-dimensional theory of gravity with negative 
cosmological constant coupled to vector field, whose general action has the
form
\begin{eqnarray}
  \label{actiune}
  I 
  \!&=&\!-\frac{1}{K_5^2}\int_{M} d^{5}x 
  \sqrt{-g}[ R - 2\Lambda-F_{\mu\nu}F^{\mu\nu} + \alpha' L_{GB}]
 \end{eqnarray}
where $F_{\mu\nu}=\partial_{\mu}A_\nu-\partial_{\nu}A_\mu$ is the gauge field, $\Lambda$ is the cosmological constant, 
and $K_5^2=16\pi G$. We use Gaussian units so that factors of $4\pi$ in the gauge 
fields can be avoided. The GB term 
$L_{GB}=R^2-4R_{\mu\nu}R^{\mu\nu}+R_{\mu\nu\alpha\beta}R^{\mu\nu\alpha\beta}$
appears in the low-energy effective {\it bosonic} string theory --- in type IIB 
superstring the leading corrections are cubic in $\alpha'$. 

Within this theory there is a straightforward generalization of the Reissner-Nordstrom (RN)
solution 
\beq
ds^2=-N(r)dt^2+N(r)^{-1}dr^2+r^2d\Sigma_3^2
\eeq
with 
\beq
\label{exact}
N(r)=k+\frac{r^2}{4\alpha'}\left[ 1+ \epsilon \sqrt{1+8\alpha'\left( \frac{m}{r^4}-\frac{1}{L^2}-\frac{q^2}{r^6}\right)}\, \right]\, ,
~~~A_{\mu}=(-\frac{\sqrt{3}q}{r^2}+\Phi)\delta_{\mu t}
\eeq
 where $\Phi$ is a constant which is chosen such that $A_{t}(r_h)=0$ and 
 $r_h$ is the largest positive root of $N(r)$ that is typically associated to 
 the outer horizon of a black hole --- note that the condition $N'(r_h)>0$ implies 
 the existence of a minimal allowed value of $r_h$. Here, $L$ is the radius of 
 $AdS$ spacetime and it is related to the cosmological constant by $\Lambda=-6/L^2$ 
 and $k={1,0,-1}$ corresponds to black holes with spherical, planar, and hyperbolic 
 horizon topologies. The expression of $N(r)$ has an extra parameter $\epsilon=\pm 1$,
 that implies the existence of two branches of solutions.

Let us discuss now some known limits of the solutions (\ref{exact}) --- more 
details can be find in \cite{torii-maeda}. 
The minus-branch solution reduces in the limit of
$\alpha'\to 0$~to the RN solution of the Einstein-Maxwell-$\Lambda$ 
system, $i.e.$ $N(r)=k-m/r^2+q^2/r^4$.
On the other hand, $N$ diverges for the $\epsilon=+1$ branch
and so there is no smooth limit in this case, since  
$N(r)= r^2/(2\alpha')+k-m/r^2+q^2/r^4-r^2/L^2$ as $\alpha'\to 0$.

The background approached asymptotically by these solutions corresponds to 
an $AdS_5$ spacetime with an effective radius
\beq
\label{Leff}
L_{eff}=L \sqrt{\frac{1+\epsilon U}{2} },~~
{\rm where ~~~}
U=\sqrt{1-\frac{8\alpha'}{L^2}}   
\eeq
This limit ($m=q=0$) corresponds to $AdS$ with higher derivatives. 
This effective radius of $AdS$ with higher derivative corrections will play an 
important role in the subsection \ref{count} where we will define the 
counterterms for the action and the stress-energy regularization. It is clear 
that by adding higher derivative corrections (even for small $\alpha'$) the 
theory contains new solutions (in our case a new branch) unavailable in general 
relativity.


\subsubsection{The counterterm method}
\label{count}
We start by reviewing some known useful facts about the quasilocal 
formalism of Brown and York \cite{Brown:1992br}. 
The gravitational field (the metric tensor) couples to the energy 
momentum-tensor (or stress tensor) of every other field in nature.
In general relativity, mass is merely one aspect of the stress tensor, and 
gravitational energy is non-local as follows from the equivalence principle. 
By choosing a coordinate system that is inertial in a given volume element 
one can make the stress-tensor vanishing (since $\Gamma^i_{kl}$ is vanishing). 
Thus, it has no meaning to speak of a definite localization of the energy of 
the gravitational field in space. One can measure the gravitational field by 
the geodesic deviation of two observers --- a single observer cannot distinguish 
it from kinematical effects. In other words, curvature cannot be measured on a 
point line, but requires a 2-surface at least. Since an appropriate definition 
of the gravitational energy cannot be found locally, a quasilocal definition 
is sought.

One way to compute the energy of a gravitational system is by enclosing it with 
a surface --- the observers living on this surface can make measurements and 
compare the results. In the quasilocal formalism, the {\it surface} stress tensor 
for spacetime and matter is defined by
\beq
T^{ab}\equiv \frac{2}{\sqrt{\gamma}}\frac{\delta S_{cl}}{\delta \gamma_{ab}}
\eeq 
where $\gamma_{ab}$ is the induced metric on the enclosing surface.
Even if there are similarities between the definiton of the matter stress tensor 
and the boundary stress tensor, it is worth to emphasize that $T^{ab}$ 
characterizes the entire system, including contributions from both the 
gravitational field and the matter fields \cite{Brown:1992br}.

As usual in gravity theories, the action (\ref{actiune}) 
should be supplemented with suitable boundary terms to obtain a 
well-defined variational principle. For Einstein gravity, one considers the
Gibbons-Hawking surface term \cite{Gib}
\begin{equation}
I_{b}^{(E)}=-\frac{1}{8\pi G}\int_{\partial \mathcal{M}}d^{4}x\sqrt{-\gamma }K~
\label{IGH}
\end{equation}
where $\gamma _{\mu \nu }$ and $K$ are the induced metric and the
trace of the extrinsic curvature of the boundary, respectively.

A similar term occurs for Gauss-Bonnet gravity
and reads \cite{Myers:1987yn} 
\begin{equation}
I_{b}^{(GB)}=-\frac{1}{8\pi G}\int_{\partial \mathcal{M}}d^{4}x\sqrt{-\gamma }%
\left\{ 2\alpha'\left( J-2E_{ab}^{(1)}K^{ab}\right) \right\}~
\label{IGB}
\end{equation}
where $ E_{ab}^{(1)}$ is the 
four-dimensional Einstein tensor 
of the metric $\gamma _{ab}$ and $J$ is the
trace of
\begin{equation}
J_{ab}=\frac{1}{3}%
(2KK_{ac}K_{b}^{c}+K_{cd}K^{cd}K_{ab}-2K_{ac}K^{cd}K_{db}-K^{2}K_{ab})
\label{Jab}
\end{equation}
Variation of the action $I+I_{b}^{(E)}+I_{b}^{(GB)}$
now gives an expression that does not contain normal derivatives of $\delta g_{ab}$.

It is well known that the total action contains divergences even 
at tree-level --- they arise from integrating over the infinite 
volume of spacetime. We regularize the divergences by using the 
procedure proposed in \cite{Balasubramanian:1999re}. 
This technique was inspired by the AdS/CFT duality and consists in adding 
suitable counterterms $I_{ct}$ to the action of the theory in order to 
ensure its finiteness.  

We have found that the action of the solutions in this paper can be regularized 
by the following counterterm 
\begin{eqnarray}
I_{\mathrm{ct}} &=&\frac{1}{8\pi G}\int_{\partial \mathcal{M}} d^{4}x\sqrt{-\gamma }
(
c_1-\frac{c_2}{2} \mathsf{R}
)
\label{Ict}
\end{eqnarray}
where $\mathsf{R}$ is the curvature scalar associated with the induced metric $\gamma $.
The consistency of the procedure requires 
\begin{eqnarray}
c_1= -\frac{1}{L_{eff}}(2+\epsilon U)~~
c_2= \frac{L_{eff}}{2}(2-\epsilon U)
\end{eqnarray} 
This counterterm is general and can be used to regularize  
the action of both branches.\footnote{A discussion of the counterterm 
method for GB gravity with cosmological constant also appears in the 
forth-coming paper \cite{eugen1}.} 
For solutions with a well defined Einstein gravity limit, one finds as  
$\alpha'\to 0$, that $c_1\to -3/L+\alpha'/L^3+O(\alpha')^2,$ $c_2 \to L/2+3\alpha'/2L+O(\alpha')^2$
that match the results in \cite{Brihaye:2008kh, Cardoso:2008gm} (see, also, \cite{Cvetic:2001bk}).

Gravitational thermodynamics is then formulated via the Euclidean path integral, where 
one integrates over all metrics and matter fields between some given initial and final 
Euclidean hypersurfaces. Semiclassically the total action is evaluated from the classical 
solution to the field equations. The thermodynamical system has a constant temperature
\beq
T_H=\frac{1}{\beta}=\frac{N'(r_h)}{4\pi}
\eeq
where $\beta$ is the periodicity of the Euclidean time determined by requiring the Euclidean 
section be free of conical singularities.

To evaluate the action, one express the bulk action as a total derivative
\begin{eqnarray}
 \frac{1}{2}(R-2\Lambda+ \alpha' L_{GB}-F^2)=
\frac{1}{r^3}
\left(-\frac{1}{2}r^3\,N'-\frac{3\,q^2}{r^2} +  {6\,r\,\alpha' \,\ (N - k) \,N' }
\right)' 
\end{eqnarray}
where a prime for a metric function
denotes a derivative with respect to the radial coordinate $r$.
After adding the boundary terms, one finds that the Euclidean action 
is finite and contains two terms, $I=I^{as}+I^{eh}$. These two terms 
represent the contributions from the boundary and the event horizon and 
their expressions are 
\begin{eqnarray}
\label{Iterm1}
&&I^{as}=\frac{3V_k\beta}{16 \pi G}m+I^{as}_0~~{\rm with~~}
I^{as}_0= k^2\frac{3\beta L^2_{eff}V_k}{64\pi G}( 3 \epsilon U-2)
\\
\label{Iterm2}
&&I^{eh}= \frac{\beta V_k }{8 \pi G}
\left(
 \frac{1}{2}(r_h^2+12k\alpha')N'(r_h)
+\frac{ 3 q^2}{r_h^2}
\right)~
\end{eqnarray}
with $V_k$ the area of the surface $\Sigma_k$.
One can easily verify that the action computed according to a background
subtraction coincides with the above expression up to the Casimir term 
$I^{as}_0$ (the background choice in this case corresponds to a $q=0$ 
vacuum EGB-AdS solution).

Varying the total action (that contains the  boundary terms 
(\ref{IGH}),(\ref{IGB}), and (\ref{Ict})) with respect to the
boundary metric $h_{ab}$, we compute the divergence-free boundary 
stress-tensor
\begin{eqnarray}
\label{bst}
T_{ab}=\frac{1}{8 \pi G}
\left(
K_{ab}-K\gamma_{ab}
+c_1\gamma_{ab}+c_2{\rm G}_{ab}
+ \frac{{\alpha}}{2} (Q_{ab}-\frac{1}{3}Q\gamma_{ab}) 
\right)~
\end{eqnarray} 
where 
\begin{eqnarray}
Q_{ab}= 
2KK_{ac}K^c_b-2 K_{ac}K^{cd}K_{db}+K_{ab}(K_{cd}K^{cd}-K^2)
\\
\nonumber
+2K \mathsf{R}_{ab}+\mathsf{R}K_{ab}
-2K^{cd}\mathsf{ R}_{cadb}-4 \mathsf{R}_{ac}K^c_b~
\end{eqnarray}
with $\mathsf{R}_{abcd}$ and $\mathsf{R}_{ab}$  denoting
the Riemann   and Ricci tensors of the boundary metric. 

Provided the boundary geometry has an isometry generated by a
Killing vector $\xi ^{i}$, a conserved charge
\beqa
{\frak Q}_{\xi }=\oint_{\Sigma }d^{3}S^{i}~\xi^{j}T_{ij}
\label{charge}
\eeqa
can be associated with a closed surface $\Sigma $ \cite{Balasubramanian:1999re}. 
Physically, this means that a collection of observers on
the hypersurface whose metric is $h_{ij}$ all observe the same value
of ${\frak Q}_{\xi }$ provided this surface has an isometry
generated by $\xi$. The mass/energy $M$
is the conserved charge associated with the Killing vector
$
\xi =\partial /\partial t$.
For charged black holes, the  expression of the nonvanishing
components of the boundary stress tensor are
\begin{eqnarray}
&&8\pi G T_w^w=\left(\frac{1}{2}m L_{eff}- k^2\frac{L^3_{eff}}{8}
(2-3\epsilon U)\right)\frac{1}{r^4}+O(1/r^6)~~
\\
&&8\pi G T_t^t=\left(-\frac{3}{2}mL_{eff}+3k^2\frac{L^3_{eff}}{8}
(2-3\epsilon U)\right)\frac{1}{r^4}+O(1/r^6)~
\end{eqnarray}
where $w$ denotes an angular direction on $\Sigma_3$
(note that in $1/r^4$ order, this is a traceless stress tensor).

Due to its high degree of symmetry, AdS space has a simple form in a 
large number of coordinate systems. By choosing different foliations of 
the spacetime one can describe boundaries that have different topologies 
and geometries (metrics), affording study of the CFT on different backgrounds. 
Specifically, we found additional Casimir-type contributions to the total energy 
depending on the slicing topology in accord with the expectations from quantum 
field theory in curved space.
This can be seen for the solutions discussed in this section, whose mass
 computed according to (\ref{charge}) is 
\begin{eqnarray}
M=\frac{3V_k}{16 \pi G}m+k^2\frac{3L_{eff}^2}{64\pi G}V_k( 3 \epsilon U-2)~
\end{eqnarray} 
where the last term is the Casimir energy. 

The metric on which the boundary 
CFT is defined is found by getting rid of the divergent conformal factor,
$h_{ab}=\lim_{r \rightarrow \infty} \frac{L_{eff}^2}{r^2}\gamma_{ab}$,
and corresponds to 
\begin{eqnarray}
\label{b-metric}
h_{ab}dx^a dx^b=-dt^2+L_{eff}^2d\Sigma^2_3
\end{eqnarray}
If such a CFT exists the theory lies in the landscape of string theory and the bulk 
theory is manifestly consistent as an effective theory, otherwise the theory is part 
of the swampland \cite{Vafa:2005ui}.

\subsubsection{Wald formalism}

One way of understanding black hole entropy comes from the use of Euclidean 
analog of a black hole spacetime. Whenever it is not possible to foliate the 
Euclidean section of a given (stationary) spacetime by a family of surfaces 
of constant time, gravitational entropy will emerge. Another approach to 
gravitational entropy is the Noether charge formalism of Wald. The relation 
between the two methods was explored in \cite{Iyer:1995kg} (see, also, 
\cite{Garfinkle:2000ms, Dutta:2006vs}).

When we add $R^2$ corrections to the action the entropy is no longer given by 
the area law --- instead, to computing the entropy of the black 
holes (\ref{exact}), we will use a more general formula proposed 
by Wald \cite{Wald:1993nt}:
\be
S=-2 \pi \int_{\cal H} d^3x {\partial L \over \partial
  R_{abcd}}\epsilon_{ab} \epsilon_{cd}
\ee
where ${\cal H}$ is the bifurcate horizon and $\epsilon_{\mu\nu}$ is
the binormal to the bifurcation surface. Interestingly enough, the entropy 
can still be expressed as a local functional evaluated at the (bifurcate) 
horizon. In this construction, the entropy was obtained from the Noether 
charge that is the integral of a $3$-form associated with the diffeomorphism 
invariance of the theory. It is worth noticing that Wald formalism can be 
applied to {\it non-extremal} black hole solutions in generally covariant 
theories of gravity.  

The most general formula for the entropy for a Lagrangean of the form 
\be
I=\int d^5x \sqrt{-g} \lf {R \over 16 \pi G}-2 \Lambda +\alpha R^2 + \beta
 R_{\mu \nu}R^{\mu \nu} + \gamma R_{\mu\nu\rho\sigma} R^{\mu\nu\rho\sigma} \rf
\ee
is given by (see apendix)
\be 
S={1 \over 4G}\int_{\cal H} d^3x \sqrt{h}\lf 1+2 K_5 \alpha R+ K_5
\beta(R-h^{ij}R_{ij} )+ 2 K_5 \gamma (R-2 h^{ij}R_{ij}+
h^{ij}h^{kl}R_{ikjl}) \rf 
\ee
We are interested in GB term for which the expression for entropy becomes,
\be
S={1 \over 4G}\int_{\cal H} d^3x \sqrt{h} \lf 1 + 2 \alpha' h^{i,k} h^{j,l} R_{ijkl} \rf
\ee
where 
\bea
\alpha ={\alpha' \over K_5} ,~~
\beta =-{4 \alpha' \over K_5},~~
\gamma  ={\alpha' \over K_5}
\eea
for GB term and $K_5 = 16 \pi G$.
It is easy to show that
\be
h^{ik} h^{jl} R_{ijkl} = {6 \over r_+^2}
\ee
and hence entropy becomes
\be
S=\frac{V_k}{4G} r_h (r_h^2+12 k\alpha')~
\ee
that matches with (\ref{entrogb}) obtained by the counterterm method.

\subsection{The grand canonical and canonical ensembles}
The results above make possible a discussion of the
thermodynamic properties
of these charged black hole solutions.  
In a very basic sense, gravitational entropy can be regarded as
arising from the Gibbs-Duhem relation applied to the path-integral
formulation of quantum gravity, which in the
semiclassical limit yields a relationship between gravitational
entropy and other relevant thermodynamic quantities. 
In this approach, the expression of the entropy is
\begin{equation}
S=\beta (M-\mu _{i}{\frak C}_{i})-I  \label{GibbsDuhem}
\end{equation}%
upon application of the Gibbs-Duhem relation to the partition 
function, with chemical potentials ${\frak C}_{i}$ and
conserved charges $\mu _{i}$.
For the situation in this work, ${\frak C}$ corresponds to the electrostatic potential
$\Phi$, while $\mu$ is the electric charge $Q$, with 
\begin{eqnarray}
\Phi=\frac{3q}{r_h^2},~~Q=\frac{V_k}{8\pi G}2\sqrt{3}q~
\end{eqnarray}
To compute the entropy, it is convenient to express everything in terms of $(r_h,q)$ 
\begin{eqnarray}
 T_H=\frac{r_h}{2\pi L^2}\frac{2r_h^2/L^2-q^2/r_h^4+k}{r_h^2/L^2+4k\alpha'/L^2},
~~
m=\frac{q^2}{r_h^2}+k r_h^2+\frac{r_h^4}{L^2}+2\alpha' k^2
\end{eqnarray} 
the action being given by the sum of (\ref{Iterm1}) and (\ref{Iterm2}).
In this way, one finds the following expression for the black hole entropy:
\begin{eqnarray}\label{entrogb}
S=\frac{V_k}{4G} r_h (r_h^2+12 k \alpha')~
\end{eqnarray} 
One can easily verify that the first law of thermodynamics
$dM=T_H dS+\Phi dQ$ also holds.

The corresponding equation of state
(analogous to $f(p,V,T)$, for, say, a gas at pressure $p$ and volume $V$)
reads
\begin{eqnarray}
\label{eqt}
T_H=\frac{1}{6\pi L^2}\sqrt{\frac{Q}{\Phi}}\frac{3Q+2\Phi(3k-4\Phi^2)L^2}
{Q+16k\alpha' \Phi }.
\end{eqnarray}
A discussion of the corresponding thermodynamical properties can also be approached.
In a grand canonical ensemble one finds the Gibbs free energy
\begin{eqnarray}
\nonumber
&&W[T_H,\Phi]=M-T_HS-Q \Phi=W_0+W_1,~
{\rm where} ~W_0=\frac{V_k}{8\pi G} \frac{3k^2}{16} 
\left(L(L+\sqrt{L^2-8\alpha'})-8\alpha'\right),
\\
&&W_1=-\frac{V_k}{8\pi G}
\frac{Q^2(3Q+4\Phi L^2(4\Phi^2 -3k))
+48\alpha'k\Phi Q(9Q+4\Phi(3k-4\Phi^2)L^2)}
{96L^2\Phi^2(Q+16k\alpha'\Phi)}
\end{eqnarray}
where $Q$ is given as $Q(T_H,\Phi)$ by the equation of state (\ref{eqt}).

One can consider insted a canonical ensemble,
where the temperature and electric charge are keept fixed.
The Helmholtz potential $F=M-TS$ in this case is
\begin{eqnarray}
&&F[T_H,Q]=F_0+F_1,~~F_0=\frac{V_k}{8\pi G} \frac{3k^2}{16} 
\left(L(L+\sqrt{L^2-8\alpha'})-8\alpha'\right),
\\
\nonumber
&&F_1= \frac{V_k}{8\pi G}
\frac{Q^2(-3Q+4\Phi(3k+20\Phi^2)L^2-144\alpha'k\Phi Q(3Q+4\Phi(k-4\Phi^2)L^2)}
{96L^2\Phi^2(Q+16k\alpha'\Phi)}
\end{eqnarray}
where the electrostatic potential $\Phi$ is given as a function of 
$T_H,Q$ by the equation of state (\ref{eqt}).

\subsection{Extremal case}
 
\subsubsection{Exact solution}
The non-extremal black holes have a non-zero temperature that can be 
evaluated by eliminating the conical singularity in the Euclidean section. 
Once we impose the periodicity condition, the Euclidean time circle closes 
off smoothly and the Euclidean geometry becomes a `cigar'. On the other 
hand, an extremal Euclidean black hole has a different topology. That is 
an infinite long throat for which the Euclidean time circle does not close 
off. In this case, one is forced to work with an arbitrary periodicity 
of the Euclidean time leading to ambiguos results (though, see 
\cite{Silva:2006xv}). 

However, on the Lorentzian section the picture is quite 
satisfactory: an extremal black hole is obtained by continuosly 
sending the temperature of a non-extremal black hole to zero. While 
the temperature vanishes, the area of the horizon can remain finite.

Thus, to obtain the extremal black hole solution we work on the Lorentzian 
section. The extremal limit can be equivalently obtained by imposing the 
constraint that the horizon is degenerate (i.e., $N(r)$ has a double root: 
$N(r_H)=N'(r_H)=0$). 

One can easily solve the equations system to obtain:
\beqa
\label{mq}
m &=& 2\left(k(k\alpha'+r_H^2)+\frac{3}{2}\frac{r_H^4}{L^2} \right)\\
q^2 &=& r_H^6\left(\frac{k}{r_H^2}+\frac{2}{L^2} \right)
\eeqa
where $r_H$ is the horizon radius. 

Using the method of \cite{Astefanesei:2006sy} (see section $4.2$) it is straightforward 
to show that the near horizon geometry of (\ref{exact}) is 
$AdS_2\times S^3(H^3)$ and just the radius of $AdS_2$ receives 
corrections (we will compute and provide its concrete value in 
section \ref{subefunction}). 

These geometries are interesting in their own right and provide the generalizations 
of Bertotti-Robinson geometries with GB term. These solutions are the topological 
product of two manifolds of constant curvature. They are conformally flat 
and are supported by a flux through $S^3(H^3)$:

\beq
\label{S3}
ds^2=-F(r)dt^2+\frac{dr^2}{F(r)}+R_{k}^2 d\Sigma_3^2,~~~{\rm
with}~~~F(r)=k+\frac{4 r^2(3R_k^2+k L^2)}{L^2(R_k^2+4 k \alpha')}
\eeq
and satisfy the equations of motion with a gauge field $A_t= r
\sqrt{12/L^2+6k/R_k^2}$ --- here $R_k$ is a constant and so the 
size of $\Sigma_3^2$ is fixed. Since $H^3$ is not compact, the 
solution also exists for a vanishig electric potential in this case.

\subsubsection{Entropy function}
\label{subefunction}
Wald's construction for the entropy can be used in any general coordinate 
invariant theory of gravity including those with higher derivative terms 
in the action. The black hole entropy is obtained in terms of the field 
configurations near the horizon --- all the physical fields (not just the 
curvature) and their derivatives should be regular at the bifurcation surface. 
This method is based on a Lagrangean derivation of the first law of black hole 
thermodynamics and can be used directly on the Lorentzian section (no `Euclideanization' 
is required). However, this method can be applied just to black holes 
with bifurcate Killing horizons and so not to extremal black holes.\footnote{It 
is also worth emphasizing that a black hole formed by a collapse process does 
not have a bifurcate horizon.}

This method was also extended by Sen to extremal black holes and it is reffered 
to as the entropy function formalism \cite{Sen:2005wa}.\footnote{It is known 
that the near horizon geometry of stationary extremal black holes contains an 
$AdS_2$ space \cite{Kunduri:2007vf} (see, also, \cite{Figueras:2008qh}).
 The entropy function is constructed, on an $SO(2,1)\times SO(3)$ (for static black holes) 
 or $SO(2,1)\times U(1)$ (for stationary black holes) symmetric (near horizon) 
 background, by taking the Legendre transform 
(with respect to the electric charges and angular momentum) of the 
reduced Lagrangian evaluated at the horizon.} Extremising the entropy function 
is equivalent to the equations of motion in the near horizon limit and its 
extremal value corresponds to the entropy. A discussion on the entropy function 
formalism and the Euclidean section method can be found in \cite{Dias:2007dj}.

In this section we use the entropy function formalism and compare the results with 
the ones in the previous subsection. The general metric of $AdS_2\times S^3$ can be 
written as
\begin{equation}
  ds^2=v_1(-\rho^2d\tau^2+\frac{1}{\rho^2}d\rho^2)+
  v_2d\Omega_3^2\, .
\end{equation}
The field strength ansatz is $F=ed\tau\wedge d\rho$ and, for this geometry, the GB term 
comes out to be GB$= -{24/ v_1v_2 }$. Thus, the entropy function $F(v_1, v_2, e, Q)$ 
is given by

\begin{eqnarray}
\label{Fmisto1}
  && F(v_1,v_2,e,Q)=2\pi [Qe-f(v_1, v_2, e)]\, ,\\
  \nonumber
  && f(v_1, v_2, e)=2\pi^2\left[-2v_2^{3/2}+6v_1\sqrt{v_2}+
    2\frac{v_2^{3/2}}{v_1}e^2+v_1v_2^{3/2}\left(\frac{12}{L^2}\right)-24\alpha' \sqrt{v_2}\, \right]\, .
\end{eqnarray}

The attractor equations are:

\begin{eqnarray}
  \label{atr1}
  \frac{\partial F}{\partial v_1} & = & 0\,\,\,\Rightarrow
  \,\,\,6v_1^2-2v_2e^2+v_1^2v_2\left(\frac{12}{L^2}\right)=0   \, ,\\
  \label{atr2}
  \frac{\partial F}{\partial v_2} & = & 0\,\,\,\Rightarrow \,\,\,-v_1v_2+
v_1^2+v_2e^2+\frac{v_1^2v_2}{2}\left(\frac{12}{L^2}\right)-4\alpha' v_1=0\, ,\\
  \label{atr4}
  \frac{\partial F}{\partial e} & = & 0\,\,\,\Rightarrow \,\,\,
  Q=8\pi^2\, \frac{v_2^{3/2}}{v_1}e\, .
\end{eqnarray}

Let us now discuss in detail these equations. One important observation is that 
by adding the GB term to the action just the second attractor equation is modified.
One can easily eliminate $v_1$ from the first equation by using the third one and 
so the value of $v_2$ does not change --- we obtain the following relation between 
the electric charge and the horizon radius ($v_2=r_H^2$):

\beq
\label{nicolita}
\tilde{Q}^2=\left( \frac{Q}{8\pi^2} \right)^2=3v_2^2\left(1+\frac{2v_2}{L^2} \right)
\eeq
Using the conventions from the section (\ref{section21}) and the relation between 
the physical electric charge and the charge parameter $\tilde{Q}=\sqrt{3}q$ we can see 
that this relation matches (\ref{mq}). It is worth emphasizing that just the radius 
of $AdS_2$ receives corrections:
\beq
v_1=\frac{1}{4}\frac{4\alpha'+v_2}{1+3v_2/L^2}
\eeq
Replacing $v_1(v_2)$ and $e(Q,v_2)$ back in the entropy function we obtain the 
entropy of the extremal black hole:

\beq
S_{extremal}=8\pi^3r_H^3\left(1+\frac{12\alpha'}{r_H^2}  \right)
\eeq

The entropy of extremal black hole has the same form as for the non-extremal 
black hole (\ref{entrogb}) ($V_k=2\pi^2$ and $G_N=1/16\pi$), though the radius of the 
horizon ($r_H$) is different. Not surprisingly, the form of the $\alpha'$ 
correction is the same --- both methods, Noether charge and entropy function, are 
based on a Lagrangean derivation.

In general there are two types of first-order corrections due to higher derivative
terms. The entropy/area law is modified due to the additional terms in the 
action\footnote{These terms are evaluated using the zeroth order solutions for the 
metric and the other fields} and/or the modification of the area due to the change 
of the metric on the horizon (the extra terms in the action may change the equations 
of motion). In our case, the horizon radius, $v_2$, of the extremal black hole remains 
unchanged after adding the GB term and so the entropy is changed due to the suplementary 
terms in the action.


\section{Charged black holes with scalar hair}
In this section we generalize the results of \cite{Liu:2004it} by including the 
GB term. We obtain numerical solutions\footnote{We thank Eugen Radu 
for advice in finding the numerical solutions and explaining to us the methods used 
in \cite{eugen}.} and discuss how the GB term affects their properties 
by using the counterterms proposed in the previous section.

\subsection{The model}
We consider the generalization of the RN black holes in a five-dimensional 
theory of gravity coupled to a set of scalars and vector fields, whose general 
action has the form 
\begin{eqnarray}
  \nonumber
  I[G_{\mu\nu},\phi^I,A_{\mu}^B]
  \!&=&\!-\frac{1}{16\pi G}\int_{M} d^{5}x 
  \sqrt{-g}[ R+\alpha'L_{GB}-G_{IJ}(\phi)\partial_\mu\phi^I\partial^\mu\phi^J \\
  &&-f_{AB}(\phi)F^{A}_{\mu\nu}F^{B\, \mu\nu}-V(\phi)],
  \label{actiongen}
\end{eqnarray}
where $F^A_{\mu\nu}$ with $A=(0, \cdots N)$ are the gauge fields, $\phi\equiv (\phi^I)$ 
with ($I=1, \cdots, n$) are the scalar fields, $V(\phi^i)$ is the scalar fields 
potential. 

The equations of motion for the metric, scalars, and the gauge fields are given by 
\cite{torii-maeda,Astefanesei:2007vh}
\begin{eqnarray}
R_{\mu \nu } -\frac{1}{2}Rg_{\mu \nu} + g_{\mu\nu}V(\phi) + \alpha'H_{\mu \nu}&
=& 8\pi GT^{matter}_{\mu\nu}
\label{einstein}
\end{eqnarray}
\begin{equation}
  \frac{1}{\sqrt{-g}}\partial_{\mu}(\sqrt{-g}G_{IJ}(\phi)\partial^{\mu}\phi^J)
  = \frac{1}{2}\left(\frac{\partial f_{AB}(\phi)}{\partial \phi^I} F^A _{\phantom{A}\mu\nu} 
  F^{B\, \mu\nu} + \frac{\partial G_{KM}(\phi)}{\partial \phi^I}\partial_{\mu}\phi^K\partial^{\mu}\phi^M + \frac{\partial V(\phi)}{\partial \phi^I}\right) \, \\
  \label{dilaton}
\end{equation}
\begin{equation}
  \partial_{\mu}\left[\sqrt{-g}\left(f_{AB}(\phi) F^{B\, \mu\nu}\right)  \right] =  0\, 
  \label{gaugefield}
\end{equation}
where $T^{matter}_{\mu\nu}$ is the matter stress tensor and $H_{\mu \nu}$ is given 
by \cite{torii-maeda}
\beq
H_{\mu\nu}=2(RR_{\mu\nu}-2R_{\mu\alpha}R^{\alpha}_{\nu}-
2R^{\alpha \beta}R_{\mu\alpha\nu\beta}+
R_{\mu}^{\alpha\beta\gamma}R_{\nu\alpha\beta\gamma})-\frac{1}{2}g_{\mu\nu}L_{GB}
\eeq

The Bianchi identities
for the gauge fields are $F_{\phantom{A}\,[\mu\nu;\lambda]}^{A}=0$.

 We assume that the scalar fields approache asymptotically
 constant values, $\phi_\infty^I$, which corresponds to an 
 extremum of the potential such that
$ dV/d\phi \big|_{\phi_\infty}=0$ and $V(\phi_\infty)=-{12}/{L^2}<0$,  with the expansion
\begin{eqnarray}
V(\phi)=V(\phi_\infty)+\frac{1}{2}\frac{\partial^2 V}{\partial \phi^I \partial\phi^J}\bigg|_{\phi=\phi_\infty}\phi^I\phi^J+\dots 
\end{eqnarray}
the scalar field masses  being set by $\partial^2V/{\partial\phi^I \partial\phi^J}\big|_{\phi_\infty}=\mu_{IJ}$.
 
Under these asumptions, the background of the theory is given by the solution 
\begin{eqnarray}
\label{m1}
 ds^2=-(k+\frac{r^2}{L_{eff}^2})dt^2+ \frac{dr^2}{k+\frac{r^2}{L_{eff}^2}}+r^2d\Sigma_3^2~~
\end{eqnarray}
with $\phi=\phi_\infty$, $k=\pm 1,0$ and the effective length scale is 
$L_{eff}=L \sqrt{ {1+ U}/{2} },$ with $U=\sqrt{1- {8\alpha'}/{L^2}} $ 
as in the case of EGB-$\Lambda$ theory.

Restricting to static solutions, we consider the metric ansatz
 \begin{eqnarray}
\label{metric-gen} 
ds^{2}= 
\frac{dr^2}{N(r)}+r^2d\Sigma^2_{3 }-  N(r)\sigma(r)^2 dt^{2},
\end{eqnarray}
and a purely-electric
abelian field ansatz $A^B=W^B(r)dt$,
the scalar fields being  also functions only of the radial coordinate $r$. 

The Maxwell equation implies the existence of the first integrals
\begin{eqnarray}
\label{fiem}
 W'^A=f^{AB} q_{B} {\frac{\sigma}{r^3}}
\end{eqnarray}
where $q_B$ are constants fixing the electric charges of solutions, 
$Q_B= q_B{V_k}/({2\pi G})$ and $f^{AB}$ is the inverse of $f_{AB}$.
The electric potentials are the integrals of $F_{rt}^B$, being fixed 
up to arbitrary constants $\Phi^B$ which are chosen such that $A_t^B$ 
vanish  on the event horizon.  

It is more convenient to combine the equations of motion (see \cite{Astefanesei:2007vh})  
to obtain the following equivalent system of differential equations:
 
\begin{eqnarray}
\nonumber
 &&\frac{3}{4}r (rN'+2N-2k)-3\alpha'(N-k)  N'
+\frac{1}{4}r^3  N G_{IJ} \phi'^I\phi'^J +\frac{1}{4}r^3 V(\phi) +\frac{1}{2} f^{AB}  q_Aq_B=0
\\
\label{eq-m}
&&\frac{\sigma'}{\sigma}=\frac{1}{12}\frac{r^3G_{IJ} \phi'^I\phi'^J}{r^2/4-\alpha'(N-k)}
\\
\nonumber
&&\frac{2}{r^3\sigma}(Nr^3\sigma G_{IJ} \phi^{J'})'=N\phi^{K'}\phi^{S'}\frac{\partial G_{KS}}{\partial \phi^I} +
2 \frac{\partial f_{AB}}{\partial \phi^I}f^{AC}f^{BD}\frac{q_C q_D}{r^6}
+\frac{\partial V}{\partial \phi^I}
\end{eqnarray}
 
 The first equation does not contain any second derivatives and is the Hamiltonian 
 constraint. We notice that the equations of motion can also be derived from the 
 one-dimensional reduced Lagrangean: 
 \begin{eqnarray}
 \nonumber
 L_{red}=-3r\sigma(r N'+2N-2k)+12\alpha'(N-k)\sigma N'
 -r^3\sigma N G_{IJ}\phi^{I'}\phi^{J'}-r^3 \sigma V(\phi)- \frac{2\sigma}{r^3}f^{AB} q_Aq_B
 \end{eqnarray}
We are interested in black hole solutions approaching asymptotically 
the background (\ref{m1}), that suggests to use the following form of 
the metric function, $N(r)$:  
\begin{eqnarray} 
N(r)=k-\frac{m(r)}{r^2}+\frac{r^2}{L^2_{eff}}
\end{eqnarray} 
The first equation in (\ref{eq-m}) implies that $m(r)$ satisfies the following equation
\begin{eqnarray} 
\nonumber
\frac{3}{4}\left(U m+ L_{eff}^2(U-1)\frac{m^2}{2r^4}\right)'=\frac{1}{4}r^3  N G_{IJ}\phi^{I'}\phi^{J'}+\frac{1}{4}r^3 (V(\phi)-V(\phi_\infty))+\frac{1}{2r^3}f^{AB}q_Aq_B
\end{eqnarray} 

The computation of the action and the stress tensor of these configurations 
can be done by using a similar approach to the one discussed in section 2 and 
we will not present the details here (without $\alpha'$ corrections, see 
\cite{Liu:2004it}). Similar to the case without scalars, the volume term in 
the action (\ref{actiongen}) has a total derivative structure and so it can 
be expressed in terms of the difference of two surface integrals.

The divergencies associated with the asymptotic AdS structure of the solutions
can be removed by supplementing (\ref{actiongen}) 
with the same boundary  counterterms (\ref{Ict}) as in section 2. After 
Wick rotating $t \to i\tau$ to the Euclidean section, we found that the 
action can be written in the usual `quantum statistical' form 
\begin{eqnarray}
\label{itot}
 I=\beta (M - Q_A\Phi^A) - \frac{V_k}{16\pi G}r_h(r_h^2+12\alpha'k)
\end{eqnarray}
where  $\beta$  is the periodicity of the Euclidean time. Here, M is 
the mass of the black hole and we will present its values for
some concrete examples in the next sections.

The value of $\beta$ is arbitrary for soliton solutions (that exist in 
the absence of gauge fields, e.g. \cite{Astefanesei:2003qy}) or extremal 
black holes. However, the regularity of the Euclideanized solutions as 
$r\to r_h$ imposes 
\begin{eqnarray}
\label{beta}
\beta=\frac{1}{T_H}=\frac{4\pi}{N'(r_h)\sigma(r_h)}
\end{eqnarray}
for non-extremal black hole solutions. 

The mass of these solutions can be computed within the quasilocal formalism
by using the generic relation (\ref{charge}), where the boundary stress tensor
is still given by (\ref{bst}). However, the situation is  different for theories 
with massless scalar fields and in the presence of massive scalar fields --- we 
shall discuss these cases separately. 

The explicit construction of solutions 
requires specification of the functions $G_{IJ}$ and $f_{AB}$. In what follows
we consider a model with one single scalar (i.e. $G_{IJ}=2\delta_{1I}\delta_{1J}$, $\phi^1=\phi$) and two gauge fields with modulus dependent  couplings of the form
\begin{equation}
\label{cuplarepebune}
  f_{AB}(\phi)=\delta_{AB}e^{\alpha_B \phi}
\end{equation}
Moreover, we shall restrict to solutions with a smooth Einstein gravity limit. 

\subsection{Unattractor solutions with a massless scalar field}
In their simplest version, these solutions have a constant value of the scalar potential,
 $V(\phi)=2\Lambda=-12/L^2$, which is the case considered here.
The generic solutions have a non-degenerate horizon and are easier to study.
Near the event horizon, they admit a power series expansion of the form (here we restrict to the first
terms in the series) 
\begin{eqnarray}
\label{nex1}
N(r)=f_1(r-r_h) +\dots,~ \sigma(r)=\sigma_h+\frac{2r_h^3\phi'^2(r_h)\sigma_h}{3r_h^2+4\alpha'k }(r-r_h) +\dots,
\\
\nonumber
~ \phi(r)=\phi_h-\frac{1}{2r_h^6f_1}(\alpha_1 e^{-\alpha_1\phi_h}q_1^2+\alpha_2e^{-\alpha_2\phi_h}q_2^2)(r-r_h)~~{~~}
\end{eqnarray}
where
\begin{eqnarray}
\label{nex2}
f_1=-\frac{2(- {6r_h^6}/{L^2}-3kr_h^4+e^{-\alpha_1\phi_{h}}q_1^2+e^{-\alpha_2\phi_{h}}q_2^2)}{3r_h^3(r_h^2+4\alpha'k)}
\end{eqnarray} 
The coefficients of all higher order terms in the expression of $N,\sigma,\phi$ 
are fixed by the two parameters $\phi_h,\sigma_h$. One can easily see that the 
condition $f_1>0$ imposes the existence of a minimal value of $r_h$ for given 
values of $\phi_h,q_1,q_2$.  

One can also construct an approximate solution at the boundary in terms of three
free parameters $\phi_{\infty}$, $\Sigma$, and $M_0$ 
\begin{equation}
\label{nex3}
N(r)=1-\frac{M_0}{r^2}+\frac{r^2}{L_{eff}^2}+\frac{e^{-\alpha_1\phi_{\infty}}q_1^2+e^{-\alpha_2\phi_{\infty}}q_2^2}{3U r^4}+\dots,~~
\end{equation}
\beqa
\phi(r)=\phi_{\infty}+\frac{\Sigma}{r^4}+\dots,~~\sigma(r)=1-\frac{4\Sigma^2}{3Ur^8}+\dots
\eeqa

The next leading term at the boundary corresponds to 
a normalizable mode and the black hole is a state in the boundary CFT.

By applying the quasilocal formalism discussed above, one finds the mass of 
these solutions
\begin{eqnarray}
M=\frac{ V_k}{8 \pi G} M_0 +  k^2\frac{3L_{eff}^2}{64\pi G}V_k( 3 U-2)
\end{eqnarray}
while the entropy of solutions has the same form as in the RN case,
$ 
S=\frac{V_k}{4G} r_h (r_h^2+12  \alpha'k)
$

We will explictly check by our numerical analysis that $M, r_h,$ and $\phi_h$ 
depend of the asymptotic boundary data ($\phi_{\infty}$). This is in contrast with 
the extremal case (see section 4) where we obtain an attractor behaviour of the 
horizon.

Although an exact solution of the equations of motion (\ref{eq-m}) 
 appears to be intractable, here we present arguments for the existence 
 of non-trivial solutions which smoothly interpolate between the 
asymptotic expansions (\ref{nex1}) and (\ref{nex3}).

Starting from the event horizon expansion (\ref{nex1}) we integrated 
the  equations towards  $r\to\infty$. The integration stops when the 
asymptotic limit (\ref{nex3}) is reached with a reasonable accuracy. 
In this approach, the input parameters are $k,r_h,q_1,q_2,\alpha_1,\alpha_2,L$ 
and the value $\phi_h$ of the scalar field on the horizon.
The equation for $\sigma$ decouples from the rest, and the requirement
that $\sigma \to 1$ as $r \to \infty$ can be relaxed during the numerical 
integration, $\sigma(r_h)$ subsequently being multiplied by an appropriate 
constant factor so that the correct asymptotic behaviour is recovered.

We have solved the equations of motion  for several values of
$\alpha_1=-\alpha_2=2a$ and a large set of  $r_h,q_1,q_2,\phi_h,L$.
We follow the usual approach and, by using a standard ordinary
differential equation solver, we evaluate  the  initial  conditions (\ref{nex1})
at $r=r_h+10^{-4}$ for global  tolerance $10^{-12}$,   for a fixed  
parameter $\phi_h$ and  integrating  towards  $r\to\infty$.

The complete classification of the solutions in the space of parameters 
is a considerable task that is not aimed in this paper. Also, we shall 
restrict to the case of spherical topology horizon, although we have found 
topological black holes as well.

\begin{figure}[ht]
\hbox to\linewidth{\hss%
\resizebox{9.6cm}{7cm}    {\includegraphics{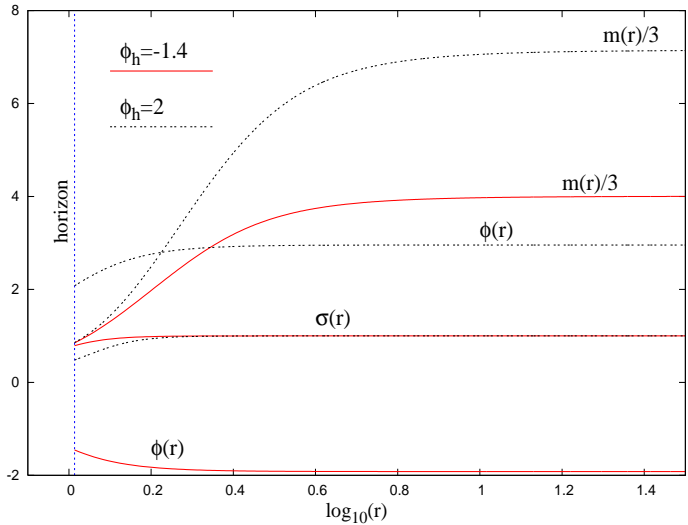}}	 
\hss}  
	\caption{ The profiles of the  functions $m(r)$, $\sigma(r)$ and 
$\phi(r)$ are shown for  typical $k=1$  non-extremal black holes with  with $r_h=1,q_1=0.3,q_2=0.5,\Lambda=-6$, $\alpha'=0.1$ and two values of
the scalar field on the event horizon }.  
\label{Fig1}
\end{figure}

\begin{figure}[ht]
\hbox to\linewidth{\hss%
\resizebox{9.6cm}{7cm}    {\includegraphics{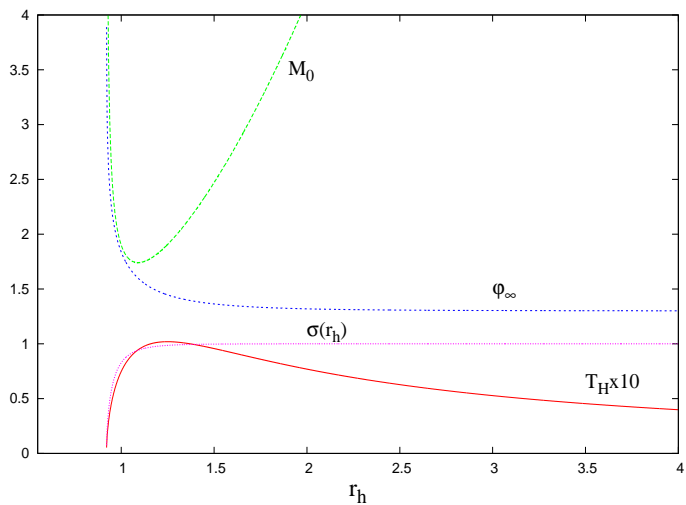}}	 
\hss}  
	\caption{The relevant parameters are ploted as a function of the 
	even horizon radius for $k=1$ non-extremal black holes with 
	$\Lambda=-0.001$, $\alpha'=0.01$, $q_1=0.3$, $q_2=0.5$,  $\alpha_1=-\alpha_2=2$.  }
\label{Fig1}
\end{figure} 

For all configurations we have studied, the metric functions $N(r)$, $\sigma(r)$, 
and the scalar $\phi(r)$ interpolate monotonically between the corresponding values 
at $r=r_h$ and the asymptotic values at infinity, without presenting 
any local extrema. In Figure 1 we plot the profiles of two typical solutions 
with different values of $\phi(r_h)$. The evolution of the solution data as a 
function of the event horizon radius is reported on Figure 2. For small values 
of $r_h$ the numerical analysis strongly suggests that an extremal black hole 
solution is approached  for a critical non-zero value of the event horizon radius.

\subsection{Configurations with a massive scalar field}

 The situation is different when allowing the scalar field to present a mass term. 
 As discussed in the last years by various authors, the field equations (\ref{eq-m}) 
 with $\alpha'=0$ admits a variety of solutions. This includes also configurations 
 where the asymptotic behaviour of the scalar field is assumed to be slower than 
 that of a localized distribution of matter.
 By relaxing the standard asymptotic conditions for asymptotically AdS solutions, 
 it is possible to preserve the original symmetries at infinity, while the conserved 
 charges are modified by including matter field terms (see $e.g.$ \cite{Henneaux:2006hk} 
 and the references there).

It would be interesting to see how these features are affected by the presence of 
a GB term in the action.
In this context, we start by discussing the issue of Breitenlohner-Freedman (BF) 
bound for a scalar field in the presence of a GB term in the 
Lagrangean. One can easily see that for $\alpha'>0$, the mass bound increases 
according to $\mu_{BF}^2=-4/L_{eff}^2$.
For example, the generic asymptotic behaviour of the scalar field $\phi(r)$ in 
the background (\ref{m1}) is  
\begin{eqnarray}
\label{d1}
\phi(r)=\frac{\phi_1}{r^{\lambda_+}}+\frac{\phi_2}{r^{\lambda_-}}
\end{eqnarray}
where $\phi_1$,  $\phi_2$ are constants and
\begin{eqnarray}
\lambda_{\pm}=2\left(1\pm\sqrt{1-\mu^2/\mu_{BF}^2}\right)
\end{eqnarray}
Here we assume $\mu^2L_{eff}^2+4 \geq 0$. Imposing that both the 
$\lambda_-$ and $\lambda_+$ solutions be normalizable results
in a supplementary conditions on the parameter $\mu^2L_{eff}^2+3<0.$
For fields that saturate the BF bound, $\lambda_+=\lambda_-$ and the 
solution is $\phi(r)={\phi_1}/{r^{\lambda}}+{\phi_2\log r}/{r^{\lambda}}$.

We shall restrict here to a scalar field $\phi$ that is tachyonic 
$(\mu^2<0)$ and its mass is in the BF range, i.e. 
decaying at infinity according to (\ref{d1}) and not saturating the 
BF bound. We also consider solutions with a well defined Einstein 
gravity limit.\footnote{A detailed discussion 
on the asymptotic form of the metric and the implications for the 
no-hair theorem can be find in \cite{Hertog:2006rr}. In our case, the only 
difference is that the solutions approach at the boundary an $AdS$ space 
with a modified radius, $L_{eff}$ given by (\ref{Leff}).}

In this case, the conserved charges are well defined and finite despite 
the fact that the scalar field falls off slower than usual. The analysis 
in the procedure of holographic renormalization is based on finding the
most general asymptotic solution of the field equations \cite{de Haro:2000xn} 
and so the fall off of the matter fields is not an input in the 
computation.\footnote{We thank Kostas Skenderis 
for discussions on holographic renormalization method for this case.}
Considering now the question of action and total mass-energy
of these solutions, one can see that  for $\phi_2\neq 0$, due 
to the back reaction of the scalar field, the boundary counterterms 
(\ref{Ict}) are not enough to cancel all divergences in the on-shell action
of the solutions. Therefore one has to supplement $I_{ct}$ with matter 
counterterms \cite{Bianchi:2001de}. The counterterms (which include
terms beyond the purely gravitational ones) are local and the action 
(and, also,  the holographic stress energy tensor) has a contribution 
from the scalar fields.

As a result, the total action of the solutions is 
 \begin{eqnarray}
\label{ta1}
I=\beta \left[ \frac{ V_k}{8 \pi G}(\frac{3M_0}{2}+M_{\phi})+M_{casimir}-Q_1\Phi^1-Q_2\Phi^2 \right]-\frac{1}{4G}(r_h^3+12 k\alpha' r_h)V_k
\end{eqnarray}
where $M_0$ is the usual mass parameter, $M_{\phi}=\lambda_{-} \phi_1 \phi_2/L_{eff}^2$ (see, e.g., 
\cite{Hertog:2006rr}), and the Casimir contribution is
\begin{eqnarray}
M_{casimir} = k^2\frac{3L_{eff}^2}{64\pi G}V_k( 3 U-2)~
\end{eqnarray} 
Thus the mass contains a supplementary term due to the slower decay of the 
scalar field. This dependence is crucial for the stress tensor to satisfy 
the correct Ward identities and to lead to sensible results.

However, from the Gibbs-Duhem relation (\ref{GibbsDuhem}) one finds that 
the entropy of these solutions is still given by the relation (\ref{entrogb}).


\section{Extremal solutions and attractor mechanism}

The numerical extremal solutions can also be discussed by using similar methods as 
in previous section. However, the situation in this case is more involved. 
Similar to the non-extremal case, one may write an approximate form of these configurations 
near the event horizon --- for static extremal black holes the near horizon geometry 
is $AdS_2\times S^3$. Let us first investigate the near horizon geometry of these 
black holes by using the entropy function formalism. For simplicity, we are again 
considering a theory with one scalar field and two $U(1)$ (electric) gauge fields 
with the couplings given by (\ref{cuplarepebune}). The general metric of $AdS_2\times S^3$ 
can be written as
\begin{equation}
  ds^2=v_1(-\rho^2d\tau^2+\frac{1}{\rho^2}d\rho^2)+
  v_2d\Omega_3^2\ 
\end{equation}

The field strength ansatz is $F^A=e^Ad\tau\wedge d\rho$. Thus, the entropy function 
$F(v_1, v_2, e^A, q_A, \phi_h)$ is similar with the one in section 2, except that we have now 
non-trivial couplings between scalars and the U(1) fields:

\begin{eqnarray}
&& F(v_1,v_2, e^A, q_A, \phi_h)=2\pi [q_Ae^A-f(v_1, v_2, e^A, q_A, \phi_h)]\, ,\\
  \nonumber
  && f(v_1, v_2, e)=2\pi^2\left[-2v_2^{3/2}+6v_1\sqrt{v_2}+
    2\frac{v_2^{3/2}}{v_1}f_{AB}(\phi_h)e^Ae^B+v_1v_2^{3/2}\left(\frac{12}{L^2}\right)-24\alpha' \sqrt{v_2}\, \right]\, .
\end{eqnarray}

By using the same trick as in section 2 we can compute the horizon radius and the value of 
the scalar at the horizon by solving the following equations 
\begin{eqnarray}
\nonumber
\alpha_1 e^{-\alpha_1\phi_h}q_1^2+\alpha_2 e^{-\alpha_2\phi_h}q_2^2=0,
~~
e^{-\alpha_1\phi_h}q_1^2+e^{-\alpha_2\phi_h}q_2^2=  3r_h^4( k+\frac{r_h^2}{L^2})
\end{eqnarray}
in terms of $q_1,q_2$ and $\alpha_1$, $\alpha_2$.

While 
$\phi_h=\frac{1}{\alpha_2-\alpha_1}\log(-({\alpha_2q_2^2})/({\alpha_1q_2^1}))$, the expression of $r_h(q_1,q_2,L)$ is very complicated and we do not present it here --- note though that the 
relation between the horizon radius and the charges is similar with (\ref{nicolita}) where $Q^2$ is 
replaced by $q_1q_2$. The horizon value of the scalar does not depend of the boundary value 
$\phi_{\infty}$ and so the near horizon geometry is universal. Consequently, the entropy 
of the extremal black hole does not depend of the boundary values of the scalar field. 
The behaviour of the scalar field is illustrated in figure 3 and the attractor mechanism 
is a direct consequence of the extremality condition.

At this point, it is worth trying to find a whole solution interpolating between the horizon and the boundary --- the entropy function assumes the existence of such a solution but does not prove it. 

Unlike in the non-extremal case, the value $\phi_h$ of the scalar field on the horizon and the 
event horizon radius $r_h$ are not free parameters and so the horizon data 
contain two essential parameters. The leading terms in this 
expansion read 
\begin{eqnarray}
\label{ex1}
N(r)=\frac{4(kL^2+3 r_h^2)}{L^2(r_h^2+4\alpha'k)}(r-r_h)^2+\dots,~~\sigma(r)=\sigma_h+\frac{2p^2 r_h^3\sigma_h\phi_1^2}{3(r_h^2+4\alpha'k)(2p-1)}(r-r_h)^{2p-1}+\dots,~~~~{~~~~}
\\
\nonumber
 \phi(r)=\phi_h+\phi_1(r-r_h)^p~+\dots~~~~~~~~~~~~~~~~~~~~~~~~~~~~~~~~~~~~{~~~~}
\end{eqnarray}
where 
\begin{eqnarray}
\nonumber
p=\frac{1}{2}\left( -1+\sqrt{ 1-3\alpha_1\alpha_2\frac{(k-\frac{2r_h^2}{L^2}) (1+\frac{4\alpha'k}{r_h^2})}{k-\frac{3r_h^2}{L^2}}  } \right)
\end{eqnarray}

Similar to the nonextremal case, we evaluate  the  initial  conditions (\ref{ex1})
at $r=r_h+10^{-4}$ for global  tolerance $10^{-14}$, integrating  towards  $r\to\infty$.
The large $r$ expansion of the extremal solutions is still given by the expression (\ref{nex3}).
Similar to the non-extremal case, the value of $\sigma_h$ in the horizon
data is not relevant in numerics and the only  parameter is $\phi_1$.
Again, we did not notice the existence of  local extrema of the functions $N(r)$, $\sigma(r)$, $\phi(r)$.

In Figure 4 we have ploted the profiles of a typical extremal black hole with  non-zero GB 
term together with the corresponding solution with $\alpha'=0$. One can see that a non-zero 
$\alpha'$ leads to a deformation of all metric functions at all scales, which holds also for 
the dilaton $\phi$. 

\begin{figure}[ht]
\hbox to\linewidth{\hss%
\resizebox{9.6cm}{7cm}    {\includegraphics{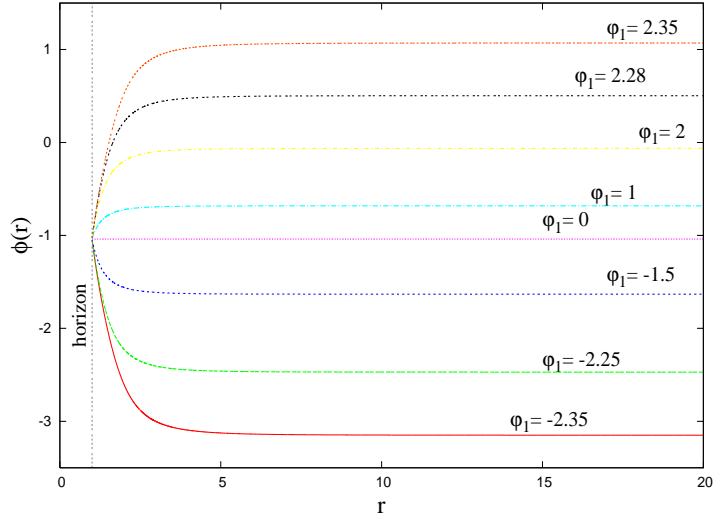}}	 
\hss}  
	\caption{ The attractor behaviour is shown for $k=1$ extremal black holes with 
	$\Lambda=-1$, $\alpha'=0.1$, $q_1=4$, $q_2=0.5$,  $\alpha_1=-\alpha_2=1/2$.  }
\label{Fig1}
\end{figure}

\begin{figure}[ht]
\hbox to\linewidth{\hss%
\resizebox{9.6cm}{7cm}    {\includegraphics{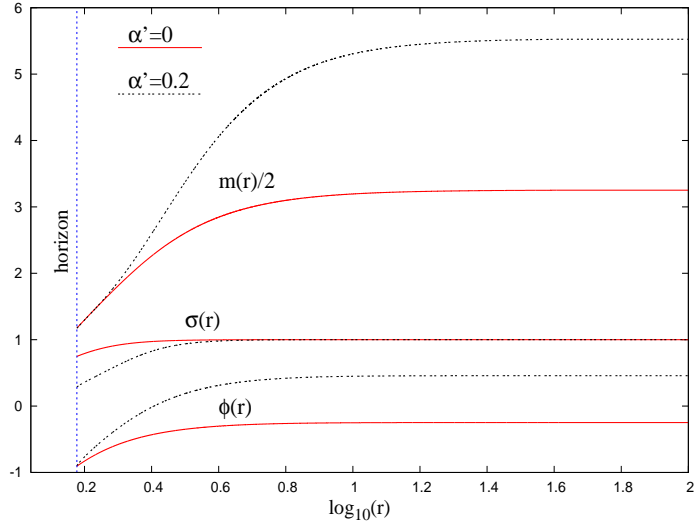}}	 
\hss}  
	\caption{ The profiles of the  functions $m(r)$, $\sigma(r)$ and 
$\phi(r)$ are shown for a typical $\alpha'=0.2$,  black hole with $k=1$, $\alpha_1=-\alpha_2=1/2,\phi_1=1,q_1=7.1,q_2=1.15$. 
For comparison, we included also the profiles  of the corresponding solution  in Einstein gravity ($\alpha'=0).$  }
\label{Fig1}
\end{figure}

It is also possible to write a simple generalization of the  
Bertotti-Robinson solution (\ref{S3}), with the same line element and the matter fields
scalar field
 \begin{eqnarray}
\phi_0=\frac{1}{\alpha_2-\alpha_1}\log(-\frac{\alpha_2q_2^2}{\alpha_1q_2^1}),~~
W^1=e^{-\alpha_1\phi_0}q_1 r,~~W^2=e^{-\alpha_2\phi_0}q_2 r,
 \end{eqnarray}
the size of  $\Sigma_3^2$ being fixed by the cosmological constant and the parameters $q_1,q_2$ as solution of  the equation
 \begin{eqnarray}
\frac{3k}{R_\Sigma^2}+\frac{6}{L^2}=e^{-\alpha_1\phi_0}q_1^2+e^{-\alpha_2\phi_0}q_2^2.
 \end{eqnarray}


\section{Discussion} 
\label{discuss}
In this paper, we have investigated the construction of black hole solutions 
in higher derivative AdS gravity. This is a self contained paper and we hope 
that our unified treatment of (non-)extremal AdS black hole solutions with GB 
term is useful to the reader. 

In the presence of higher derivative terms the area law is modified. Wald proposed 
a new formula for the themodynamical entropy such that the first law of 
black hole mechanics remains valid. In this paper, we have taken a slightly 
different route in deriving the first law and studied the thermodynamical properties 
of these black holes. The main tool that we have used in the non-extremal case is the 
counterterm method. We have explicitly constructed the counterterms that regularize the 
action and the stress tensor and applied the holographic renormalization method in 
several concrete examples\footnote{We also have checked this method for other 
solutions with non-trivial boundary topology \cite{Astefanesei:2005eq}, but we 
hope to present these results elsewhere.} --- we found perfect agreement with 
the results obtained by Wald formalism. The method, as developed in 
\cite{de Haro:2000xn}, constructs unambigously the counterterms that render 
the action finite on any solution of the field equations. In other words, the
method does not proceed with postulating a set of counterterms and
then checking that these work for a class of solutions. Rather one first 
obtains the behaviour of the on-shell action on the most general solution
and then from here one finds the set of counterterms. It would be interesting 
to understand better how the counterterms we proposed fit in this general framework.
 
AdS spacetime is geodesically complete, but the light cones flare out in such a 
way that particles can exit from the space --- and also information can come into 
the space --- within a finite time. Thus, AdS is not a globally hyperbolic spacetime 
and so the boundary conditions should play an important 
role. Indeed, within the AdS/CFT duality, various deformations of the AdS 
boundary conditions are interpreted as dual to deformations of the CFT. 
If such a CFT exists the theory lies in the landscape of string theory and the bulk 
theory is manifestly consistent as an effective theory, otherwise the theory is part 
of the swampland \cite{Vafa:2005ui}.

A `ground state' is defined as a state that extremizes the Hamiltonian over 
the class of vacuum states which all have a given boundary topology. It is well 
known that by using different foliations of AdS space one can describe boundaries 
that have different topologies affording the study of CFT on different backgrounds. 
The diffeomorphisms in the bulk are equivalent with the conformal transformations 
in the boundary, and different boundary topologies are related by {\it singular} 
conformal transformations. Therefore, for AdS gravity, the correct variational 
problem requires keeping fixed a conformal structure rather than a boundary 
metric. The variational problem for two derivative AdS gravity was discussed in 
detail in \cite{Papadimitriou:2005ii}. This paper also analyzes Wald's method for 
asymptotically locally AdS spacetimes, shows in generality that the results of this
method agree with the holographic results, and establishes the 1st law of thermodynamics.

Our results fit in this general framework. For black holes with a boundary topology 
of $R\times S^3(H^3)$ we found additional Casimir-type contributions to the energy. 
That is in accord with the expectations from quantum field theory in curved space: 
for the Casimir effect, the global structure is reflected nontrivially in the ground 
state of the quantum field. 


We have also constructed extremal solutions and investigated their properties.
Wald formalism was extended by Sen to extremal black holes. The advantage 
of this method is that the higher derivatives terms can be incorporated easily, 
but the method can not be used to determine the properties of the 
solution away from the horizon. However, in section 4, we have constructed 
numerical solutions that interpolate between the horizon and the boundary. 
Thus, we were able to safely apply the entropy function formalism 
to study their properties.\footnote{Entropy function formalism was applied to 
black holes in AdS space in \cite{Astefanesei:2007vh, Morales:2006gm}.}

Unlike the non-extremal case where the near horizon geometry (and the entropy) 
depends on the boundary values of the moduli, in the extremal case, the near 
horizon geometry is universal and is determined by only the charge parameters.
We have also constructed numerical solutions for which the near horizon geometry is 
$AdS_2\times H^3 (T^3)$, though we do not present the details here. We have found 
that, in all these cases, the scalar fields are attracted to fixed values at 
the horizon. This does not come as a surprise since it is known that the `long 
throat' of $AdS_2$ is at the basis of attractor mechanism. 

A detailed analysis of the attractor mechanism and interpretations within 
the AdS/CFT duality can be find in \cite{Astefanesei:2007vh}.\footnote{In 
this paper we are interested in AdS black holes, for more details on attractor 
mechanism in flat space one can consult the nice reviews 
\cite{Andrianopoli:2006ub} and references therein.} 
After embedding in string theory, the moduli flow becomes a holographic 
renormalization group (RG) flow. The idea that 
the IR end-point of a QFT RG flow does not depend upon UV details becomes 
in the holographic context the statement that the bulk solution in 
the near horizon limit does not depend upon the details at the 
boundary (asymptotic values of the moduli). That is a holographic 
interpretation of attractor mechanism \cite{Astefanesei:2007vh}.

When the scalars potential is not a constant, a general analysis 
of the attractor mechanism is difficult. First of all, if the 
boundary values of the moduli are fixed to a minimum of the potential 
it is not clear how `to fly' to IR horizon where the moduli may 
get different values depending of charges (the existence of extremal 
solutions in this case is problematic). However, if the potential 
has flat directions it may be possible to perturb along these directions.
Therefore, a discussion of the attractor mechanism for a non constant 
scalars potential should be made case by case.

Although the focus of this paper has been on solutions 
in higher derivative AdS gravity, it will be interesting to develop a 
similar technique for asymptotically flat solutions. In particular, 
it will be interesting to find a boundary stress tensor analogous to 
the one for two derivative gravity \cite{Astefanesei:2005ad}.

\acknowledgments 

We thank Rajesh Gopakumar and Eugen Radu for collaboration in the 
initial stages of this work and for further valuable discussions, 
Ashoke Sen for enlightening discussions, and Kostas Skenderis for 
important comments on holographic renormalization method. DA would 
also like to thank Kentaro Hanaki for interesting conversations. The 
work at HRI is supported by the Department of Atomic Energy, Government 
of India. DA would like to thank TITECH, Tokyo for hospitality during 
part of this work and NSERC of Canada for support. NB and SD acknowledge 
the hospitality of the Max Planck Institute in Potsdam during the last 
stages of this research.
\appendix

\section{Wald formalism for Gauss-Bonet action}
The action in presence of generalised Gauss-Bonnet term is of the form:
\be
I=\int d^5x \sqrt{-g} \lf {R \over 16 \pi G}-2 \Lambda - {F_{\mu \nu}F^{\mu
    \nu}\over 16 \pi G}+\alpha R^2 + \beta
 R_{\mu \nu}R^{\mu \nu} + \gamma R_{\mu\nu\rho\sigma} R^{\mu\nu\rho\sigma} \rf
\ee
Using Wald formalism, the entropy in presence of this term is given by

\be 
S={1 \over 4G}\int_{\cal H} d^3x \sqrt{h}\lf 1+2 K_5 \alpha R+ K_5
\beta(R-h^{ij}R_{ij} )+ 2 K_5 \gamma (R-2 h^{ij}R_{ij}+
h^{ij}h^{kl}R_{ikjl}) \rf 
\ee
where h is the induced metric on the boundary and $K_5=16 \pi G$.

Here, we present a detailed proof of this formula.

General Wald formula for the entropy for any Lagrangean L is 
\be
S=-2 \pi \int_{\cal H} d^3x \sqrt{h}{\partial L \over \partial
  R_{abcd}}\epsilon_{ab} \epsilon_{cd}
\ee
where ${\cal H}$ is the bifurcate horizon and $\epsilon_{\mu\nu}$ is
the binormal to the bifurcation surface, normalized such that
$\epsilon_{\mu\nu}\epsilon^{\mu\nu}=-2$.We can take 
\be
\epsilon_{\mu\nu}=\xi_\mu \eta_\nu - \xi_\nu\eta_\mu,
\ee
where $\xi$ and $\eta$ are the null vectors normal to the bifurcate 
Killing horizon, with $\xi . \eta =1$. We will take 
\be
\xi={\partial \over \partial t}
\ee
which is null at the bifurcate horizon. Then $\eta$ can be
\be
\eta= -{1 \over g_{tt}}{\partial \over \partial t}-{\partial \over \partial r}
\ee
Now with all these definitions, we can proceed to compute the entropy
using Wald formalism. We can write the Einstein-Hilbert term using its
symmetries as
\be
R={1 \over 2}(g^{ac} g^{bd}-g^{ad}g^{bc})R_{abcd}
\ee
So the leading piece in entropy is
\be
S_0=- 2 \pi \int d^x \sqrt{h}{1 \over 16 \pi G}{\partial R \over 
\partial R_{abcd}}\epsilon_{ab}\epsilon_{cd} ={A \over 4\pi}
\ee

$R^2$ part:\\
\be
{\partial R^2 \over \partial R{abcd}}\epsilon_{ab}\epsilon_{cd}=-4 R
\ee
So, the contribution to the Entropy is,=
\be
S_1= {1 \over 4 G}\int d^3 x \sqrt{h}2 K_5 \alpha R
\ee
$R^{ij}R_{ij}$ part:\\
\bea
{\partial (R^{ij}R_{ij}) \over \partial R{abcd}}\epsilon_{ab}\epsilon_{cd}&=&
2 R^{ij}g^{kl}\delta^{a}_k \delta^{b}_i \delta^{c}_l \delta^{d}_j\epsilon_{ab}
\epsilon_{cd} \nn \\
&=& 2 R^{bd}\epsilon_{ab}\epsilon^{a}_d \nn \\
&=& - 2 R^{bd}(\xi_b \eta _d + \xi _d \eta_b),
\eea
where we have used the definition of binornal. Now, using the
following relation between the induced metric and the original metric
\be
h_{bd}=g_{bd}- (\xi_b \eta _d + \xi _d \eta_b)
\ee
we get that the contribution to the entropy is
\be
S_2={1 \over 4G} \int d^3 x \sqrt{h} K_5 \beta (R-h^{ab}R_{ab})
\ee
$R^{ijkl}R_{ijkl}$ part: \\
\bea
{\partial (R^{ijkl}R_{ijkl}) \over \partial R{abcd}}\epsilon_{ab}\epsilon_{cd}
&=& 2 R^{ijkl} \delta^{a}_i \delta^{b}j \delta^{c}_k \delta^{d}_l \nn \\
&=& R^{abcd}(\xi_a \eta_b- \xi_b \eta_a)(\xi_c \eta_d - \xi_d \eta_c) \nn \\
&=& - 2 R^{abcd}(g_{ac}-h_{ac})(g_{bd}-h_{bd}) \nn \\
&=& -2 (R - 2 h^{bd}R_{bd}+ h^{ac}h^{bd}R_{abcd}).
\eea
So the contribution to the entropy is 
\be
S_3={1 \over 4 G}\int d^3 x \sqrt{h}2 K_5 \gamma (R - 2 h^{bd}R_{bd}+ 
h^{ac}h^{bd}R_{abcd})
\ee

Thus we get the net entropy due to the presence of the GB
term in the action as $S=S_0+S_1+S_2+S_3$ --- we used this expression 
in section 2.





\end{document}